\documentclass[12pt]{iopart}
\usepackage{iopams}
\usepackage{epsfig}
\usepackage{algorithm}
\usepackage{algpseudocode}
\usepackage{color}
\usepackage{amssymb}
\usepackage{amstext}

    {\end{array}\!\right)}
    {\end{array}\right.}

\begin{document}

\title[Robust quantum gate]{Learning robust control for generating universal quantum gates}

\author{Daoyi Dong}

\address{School of Information Technology
and Electrical Engineering, University of New South Wales, Canberra, ACT 2600, Australia.}
\ead{daoyidong@gmail.com}

\author{Chengzhi Wu and Chunlin Chen}

\address{Department of Control and Systems Engineering,
School of Management and Engineering
Nanjing University, Nanjing 210093, China}
\ead{clchen@nju.edu.cn}

\author{Bo Qi}

\address{Key Laboratory of Systems and Control, Academy of Mathematics and Systems Science, Chinese Academy of Sciences, Beijing 100190, China.}

\author{Ian R. Petersen}

\address{School of Information Technology
and Electrical Engineering, University of New South Wales, Canberra, ACT 2600, Australia.}

\author{Franco Nori}

\address{CEMS, RIKEN, Saitama, 351-0198, Japan.\\
Physics Department, The University of Michigan, Ann Arbor, Michigan 48109-1040, USA.}
\ead{fnori@riken.jp}

\begin{abstract}
Constructing a set of universal quantum gates is a fundamental task for quantum computation. The existence of noises, disturbances and fluctuations is unavoidable during the process of implementing quantum gates for most practical quantum systems. This paper employs a sampling-based learning method to find robust control pulses for generating a set of universal quantum gates. Numerical results show that the learned robust control fields are insensitive to disturbances, uncertainties and fluctuations during the process of realizing universal quantum gates.
\end{abstract}

Quantum information technology has witnessed rapid development in the last twenty years \cite{quantum-technology-2010}. An important task to implement quantum computation is the realization
of quantum gates. It is well known that a suitable set of single-qubit and two-qubit quantum gates can accomplish universal quantum computation. A universal gate set may consist of a quantum phase
gate (S gate), a Hadamard gate (H gate), a $\pi/8$ gate (T$_{\frac{\pi}{8}}$ gate), and a CNOT gate \cite{Nielsen-and-Chuang-2000}. Realizing such a universal gate set is a fundamental objective in quantum computation.

In
practical applications, it is inevitable that there exist
different uncertainties, inaccuracies and disturbances in external fields, or system
Hamiltonians \cite{Pravia-et-al-2003,Wu-et-al-2013,Dong-and-Petersen-2009NJP,Carignan-Dugas-et-al-2015}. Many
cases of unknown information and errors, such as imprecise Hamiltonian modeling and inaccurate control pulses, can also be treated as
uncertainties. Hence, the requirement of a certain degree of
robustness against possible uncertainties and noises has been recognized as one of the key properties for a reliable
quantum information processor. Several methods have been developed
to enhance robustness and reliability in quantum information processing \cite{James-et-al-2008,Qi-2013,Brif-et-al-2010,Theis-et-al-2016,Kimmel-et-al-2015}. Feedback control theory \cite{Wiseman-and-Milburn-2009}, including measurement-based feedback and coherent feedback \cite{James-et-al-2008}, has been developed to achieve improved performance of robustness in quantum manipulation problems. From the perspective of experimental implementation, open-loop control is usually more feasible and practical. Dynamical decoupling \cite{Viola-et-al-1999,Khodjasteh-et-al-2010,Bermudez-et-al-2012} and noise filtering \cite{noise-filtering} have been developed for enhancing robustness performance in manipulating quantum states or
quantum gates. Optimal control methods such as
sequential convex programming \cite{Kosut-et-al-2013} and gradient-based optimal algorithms (e.g., GRAPE \cite{Skinner-et-al-2003}) can also be used to
design robust control fields for manipulating quantum systems.

In this paper, we apply a learning-based open-loop
control method \cite{Chen-et-al-2013} to guide the
design of robust control fields for construction of universal quantum gates. In particular, we aim to generate the set of universal quantum gates \{S, H, T$_{\frac{\pi}{8}}$, CNOT\}.The results show that the learning control method can efficiently find optimal control fields and the designed control fields are insensitive to different fluctuations and uncertainties in the process of generating quantum gates. The quantum gates with the designed control fields can have improved robustness and reliability.

\section*{Results}
\subsection*{Optimal control results of one-qubit quantum gates.}
In this section, we consider optimal control of the one-qubit gates \{S, H, T$_{\frac{\pi}{8}}$\}. Denote the Pauli
matrices as $\sigma=(\sigma_{x},\sigma_{y},\sigma_{z})$
%
and let the free Hamiltonian be $H_{0}=\omega_{0}\sigma_{z}$, with constant $\omega_{0}$.
To construct a one-qubit quantum gate
we use the control Hamiltonian of
$H_{c}=\omega_{x}(t)\sigma_{x}$, with time-varying control $\omega_{x}(t)$.
Now we use the gradient-based learning method (see the Methods section) to construct the quantum gates H, S, and T$_{\frac{\pi}{8}}$. The index of \emph{infidelity} $\bar{F}$ (1 minus the fidelity $F$) is used to characterize the error, and atomic units are adopted. We assume $\omega_{0}=1$, $\omega_{x}(t)\in[-5, 5]$, and the terminal time $T=8$. We use piece-wise constant pulses to approximate the learned control field. We divide the time $T$ into 200 intervals, where a constant pulse is used within each interval. The initial control field is set as $\omega_x(t)=\sin t$ and the step-size is set as 0.5.

As shown in Fig. 1, the H, S and T$_{\frac{\pi}{8}}$ quantum gates can achieve a precision (fidelity) of around $1-10^{-15}$, after 70 iterations. The algorithm quickly converges and we can easily find optimal control pulses to generate the H, S and T$_{\frac{\pi}{8}}$ gates. We further consider the relationship between the infidelity $\bar{F}$ and the terminal time $T$. For example, the relationship of the infidelity $\lg\bar{F}$ (here we use the logarithm of $\bar{F}$, i.e., $\lg\bar{F}$ or $\log_{10}\bar{F}$) versus time $T$ for the H gate is shown in Fig. 2. If we fix a bound on the control amplitude, the algorithm cannot achieve good performance if the time $T$ is too short, because it may not guarantee the controllability within $T$. For a smaller bound on the control amplitude, we may need a longer terminal time $T$ to achieve the required fidelity.

\begin{figure}
\centering
\includegraphics[width=0.9\textwidth]{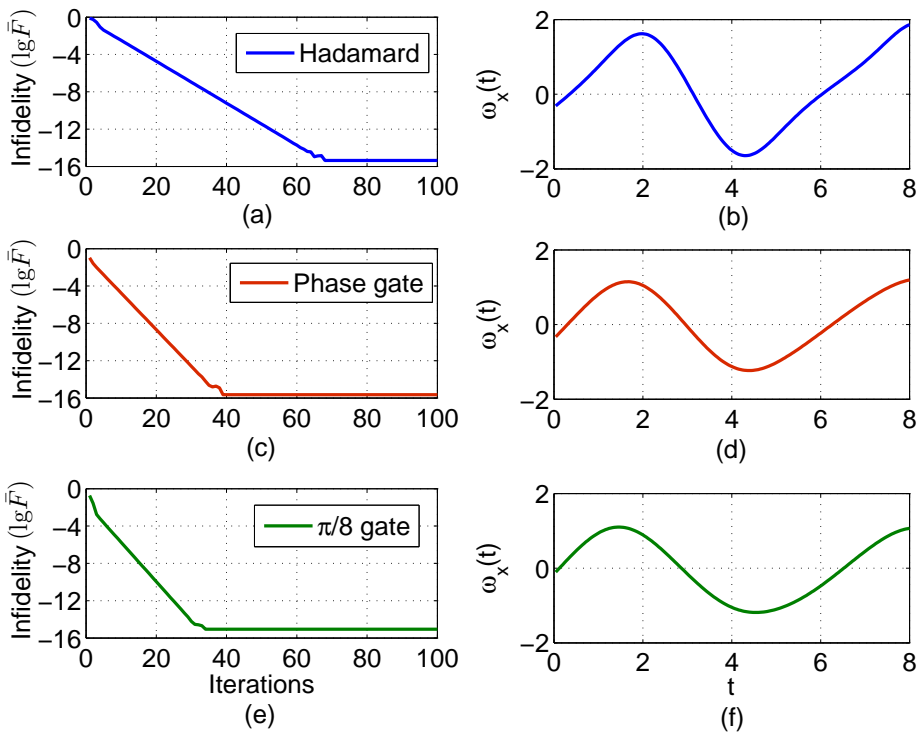}\label{optimal}
\caption{Learned optimal control for generating quantum gates H, S and T$_{\frac{\pi}{8}}$. (a), (c), (e) The infidelity ($\lg\bar{F}$) versus iterations; (b), (d), (f) The learned control fields.}
\end{figure}

\begin{figure}
\centering
\includegraphics[width=0.6\textwidth]{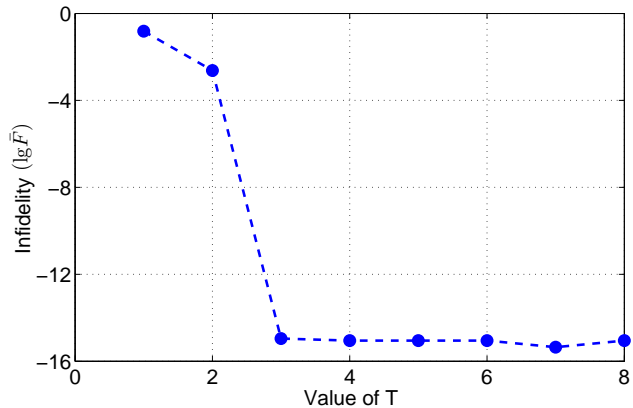}
\caption{Infidelity ($\lg\bar{F}$) versus the terminal time $T$.}\label{qubit-optimal-time}
\end{figure}

\subsection*{Robust control results of one-qubit quantum gates.}
Considering the existence of uncertainties, we assume that the Hamiltonian can be described as
\begin{equation}
H(t)=f_{0}(\epsilon_{0})\omega_{0}\sigma_{z}+f_{1}(\epsilon_{1})\omega_{x}(t)\sigma_{x}.
\end{equation}
For simplicity, we assume $f_{0}(\epsilon_{0})=\epsilon_{0}$ and $f_{1}(\epsilon_{1})=\epsilon_{1}$,
and both uncertain parameters $\epsilon_{0}$ and $\epsilon_{1}$ have uniform distributions with the same bound on the uncertainties $E=0.2$ (i.e., 40\% fluctuations, $\epsilon_{0}\in [0.8, 1.2]$ and $\epsilon_{1}\in [0.8, 1.2]$). Using the SLC method \cite{Chen-et-al-2013} (see the Methods section), an augmented system is constructed by selecting $N_{0}=5$ values for $\epsilon_{0}$, and $N_{1}=5$ values for $\epsilon_{1}$. The samples are selected as $(\epsilon_{0}, \epsilon_{1})\in \{(0.8+0.04(2m-1), 0.8+0.04(2n-1))|  m, n=1,2,\ldots,5\}$.
Fig. 3 shows the results for three classes of quantum gates: S, H and T$_{\frac{\pi}{8}}$, respectively.
After 100,000 iterations, the precision reaches 0.9979 for the H gate, 0.9976 for the S gate and 0.9991 for the T$_{\frac{\pi}{8}}$ gate, respectively. The corresponding control fields are shown in Figs. 3(b), 3(d) and 3(f). Then the learned fields are applied to 2000 additional samples that are generated randomly by selecting values of the uncertainty parameters according to a uniform distribution. The average fidelity reaches 0.9976  for the H gate, 0.9973 for the S gate, and 0.9989 for the T$_{\frac{\pi}{8}}$ gate, respectively.

In the laboratory, it may be easier for some quantum systems to generate discrete control pulses with constant amplitudes. Here, we consider the performance using different numbers of control pulses to approximate the fields. For the S gate, the relationship of the number of pulses versus the average fidelity is shown in Fig. 4. From Fig. 4, it is clear that excellent performance can be achieved even if we only use $20\sim40$ control pulses to realize the approximation of the continuous control fields. Hence, we use 40 pulses to implement the control field in the following numerical calculations.


\begin{figure}
\centering
\includegraphics[width=0.9\textwidth]{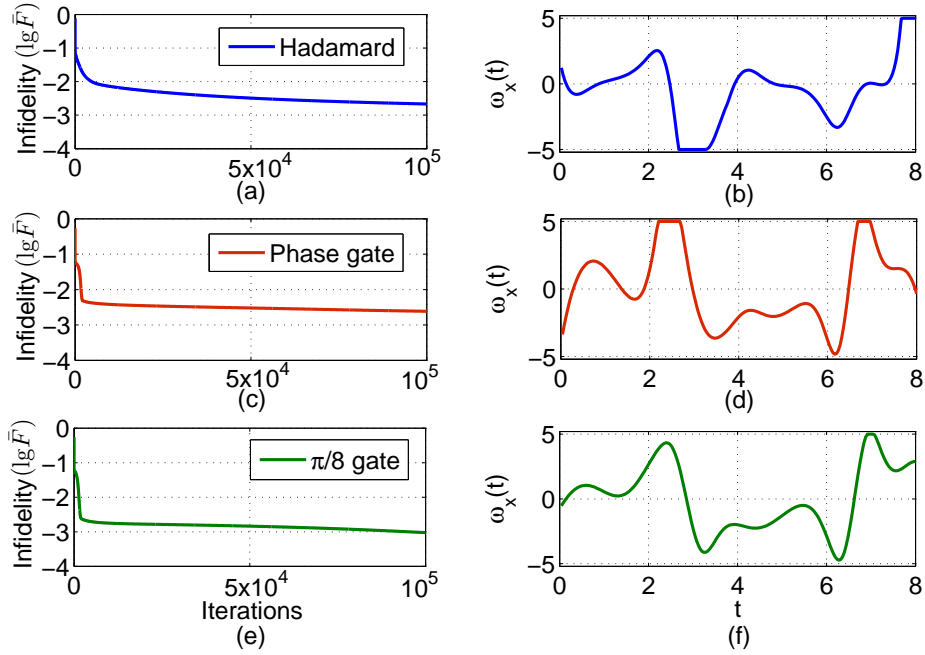}
\caption{Robust control performance for the S, H, T$_{\frac{\pi}{8}}$ gates with parameter fluctuations on $\omega_0$ and $\omega_x$. (a), (c), (e) The infidelity ($\lg\bar{F}$) versus iterations; (b), (d), (f) Learned robust control fields.}\label{qubit-omega0x}
\end{figure}

\begin{figure}
\centering
\includegraphics[width=0.6\textwidth]{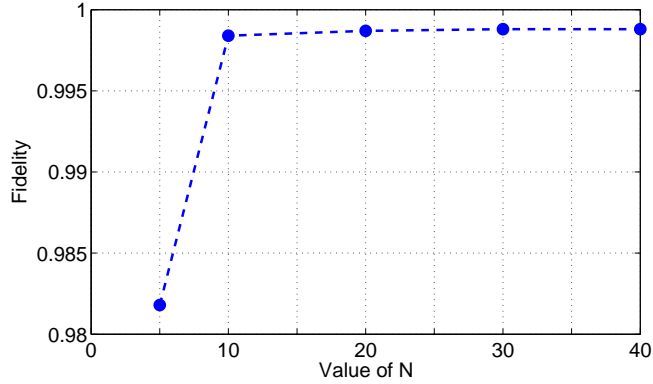}
\caption{Fidelity vs the number of sub-pulses for the S gate.}\label{qubit-intervals}
\end{figure}




We further consider the effect of the uncertainty bounds on the robustness performance. Fig. 5 shows the performance of the system when the parameter fluctuations have different bounds for the S gate. Here, we assume $E_0=E_1=E$ and $N_{0}=N_{1}=5$. Although the performance decreases when the bounds on the fluctuations increase, the control fields still can drive the system to the target gate with a high average fidelity of above 0.9950, even with $60\%$ fluctuations ($E=0.3$).

\begin{figure}
\centering
\includegraphics[width=0.6\textwidth]{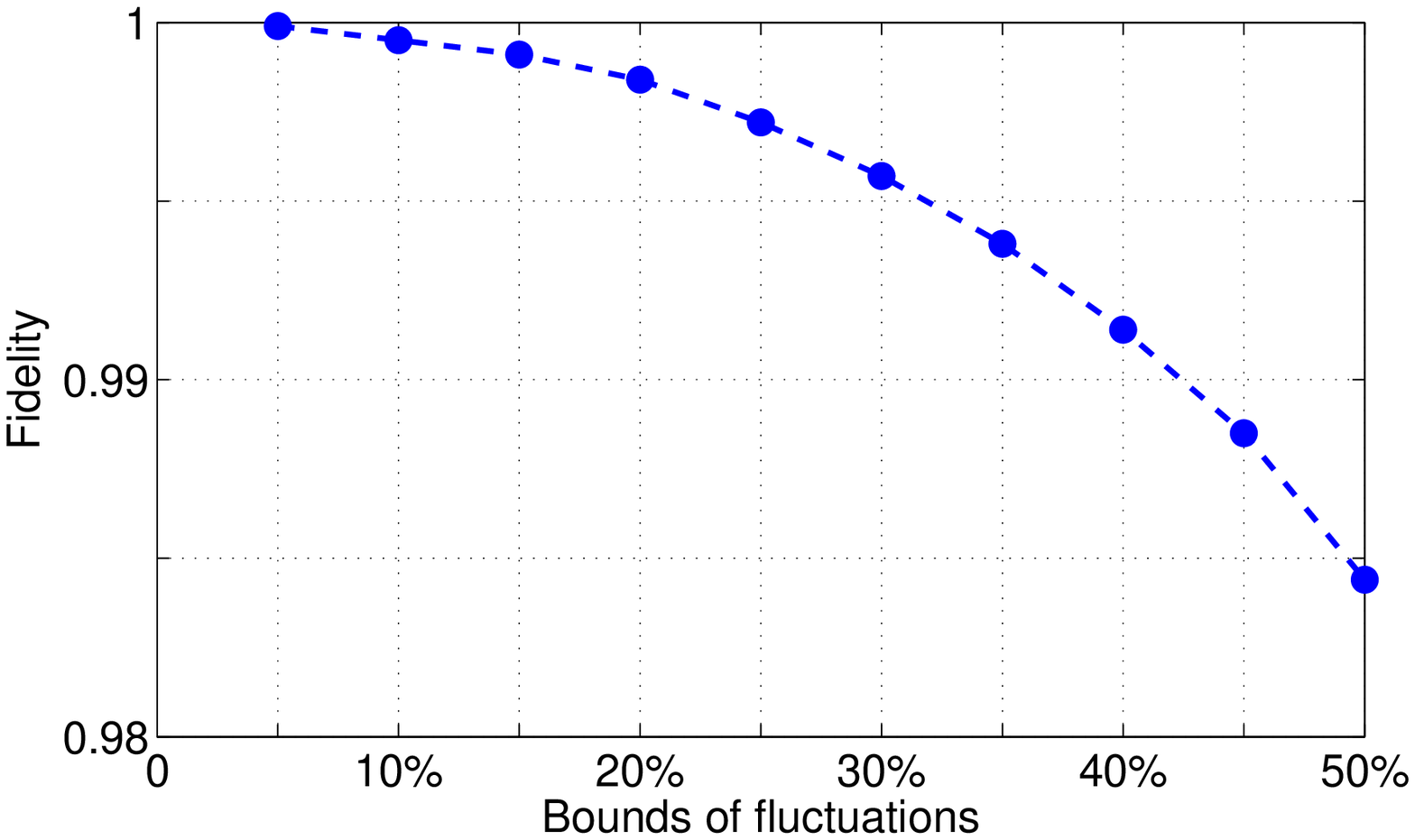}
\caption{The bounds on the fluctuations versus the average fidelity for the S gate, with parameter fluctuations on both $\omega_0$ and $\omega_x$.
}\label{H-gate-different-bound}
\end{figure}

\subsection*{Open dissipative systems for one-qubit quantum gates.}
Many quantum systems can be used to realize quantum gates. In particular, superconducting
quantum systems \cite{You-and-Nori-2011,Xiang-et-al-2013,Steffen-et-al-2010,You-et-al-2007,Yan-et-al-2016,Lv-et-al-2012,Barends-et-al-2014,Chen-et-al-2016,Friis-et-al-2015} are one of the most promising systems for the implementation of
 quantum computation due to their advantages,
such as design flexibility, tunability and scalability. For superconducting quantum circuits, it is convenient to control
the systems by adjusting external parameters such as
voltages, currents, and microwave photons, and it is also possible to turn on and off the coupling between
two qubits at will \cite{Clarke-and-Wilhelm-2008,You-and-Nori-2005}. In practical applications, the existence of fluctuations (e.g., fluctuations
in magnetic and electric fields), inaccuracies (e.g., inaccurate operation
in the coupling between qubits), and decoherence,
may degrade the performance of reliability and robustness in quantum computation \cite{Bylander-et-al-2011}. In Ref. \cite{Dong-SR}, the robustness problem for steering quantum states in superconducting quantum circuits has been investigated using the SLC method. Here, we apply the SLC method \cite{Chen-et-al-2013,Dong-SR} to design control fields that are
robust against different inaccuracies and fluctuations for implementing quantum gates. Now, we consider a flux qubit subject to decoherence to generate the S, H, and T$_{\frac{\pi}{8}}$ quantum gates. We assume that the dynamics of the flux qubit can be described as
\begin{equation}
\dot{\rho}(t)=-i[H(t), \rho(t)]+\Gamma_{1}\mathcal{D}[\sigma_{-}]\rho(t)+\Gamma_{\varphi}\mathcal{D}[\sigma_{z}]\rho(t)\equiv \mathcal{L}\rho(t)
\end{equation}
where $$\mathcal{D}[c]\rho=c\rho c^{\dagger}-\frac{1}{2}c^{\dagger} c\rho-\frac{1}{2}\rho c^{\dagger} c.$$
Here, $\Gamma_{1}$ and $\Gamma_{\varphi}$ are the relaxation rate and dephasing rate of the system, respectively. Considering the experiment \cite{Bylander-et-al-2011}, we choose $\Gamma_{1}=10^{5}\ \text{s}^{-1}$ and $\Gamma_{\varphi}=10^{6}\ \text{s}^{-1}$. Let $T=5 \ \text{ns}$ and assume that the control Hamiltonian is described as
$$H(t)=u_{x}(t)\sigma_{x}+u_{z}(t)\sigma_{z}.$$
We assume that there exist fluctuations (with the fluctuation bound 0.2) in the relaxation rate and dephasing rate. Using the OPEN GRAPE algorithm \cite{OpenGrape4-Glaser-2011} (see the Methods section), we can learn robust control fields for generating the three classes of quantum gates. The results are shown in Fig. 6. After 80 iterations, the fidelity of all three gates reaches 0.9948 using 40 control pulses for each class of quantum gates.
\begin{figure}
\centering
\includegraphics[width=1.0\textwidth]{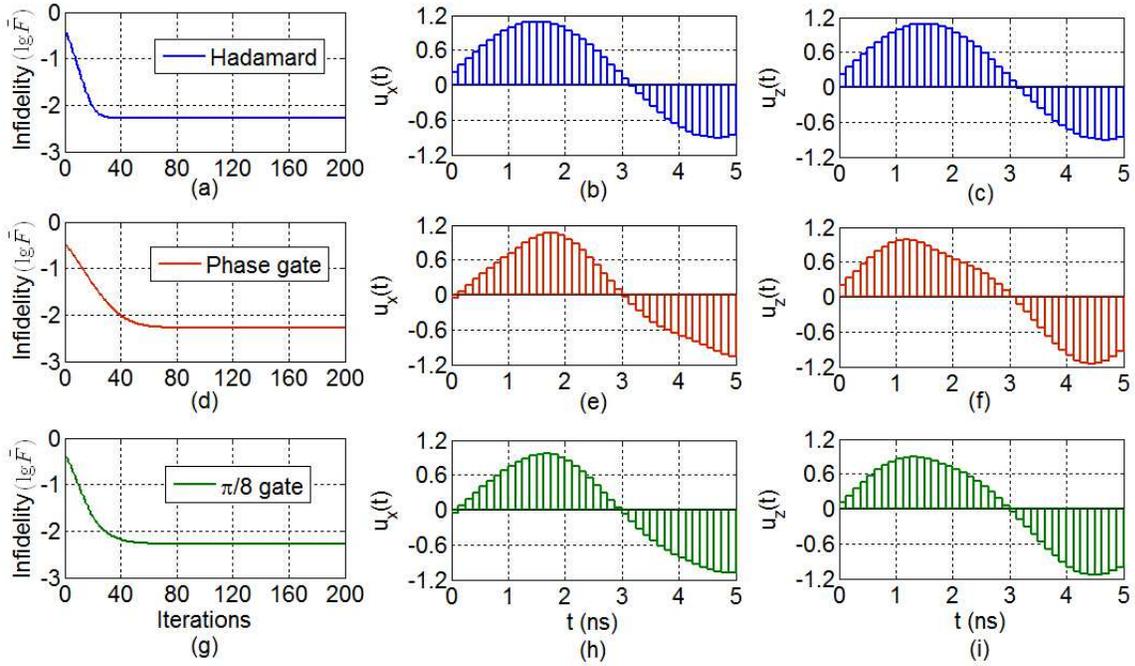}
\caption{Robust control performance for the S, H, T$_{\frac{\pi}{8}}$ gates, with parameter fluctuations in the relaxation rate and dephasing rate for open quantum systems. (a), (d), (g) Convergence for the H gate, the S gate, and the T$_{\frac{\pi}{8}}$ gate, respectively. The other sub-figures show the robust control pulses.}\label{open-qubit}
\end{figure}

\subsection*{Quantum CNOT gate.}
In this section, we consider the problem of finding robust control pulses for generating quantum CNOT gates.
In particular, we consider the example based on the two coupled superconducting phase qubits in Ref. \cite{Martinis-et-al-2011PRL}, which has also been discussed for the robust control of quantum states in \cite{Dong-SR}. Each phase qubit is a nonlinear resonator built from an Al/AlO$_x$/Al Josephson junction, and two qubits are coupled via a modular four-terminal device (for details, see Fig. 1 in Ref. \cite{Martinis-et-al-2011PRL}).
We assume that the Hamiltonian has the following form (due to possible fluctuations and uncertainties):
\begin{equation}
H=\frac{\hbar \epsilon_{1} \omega_{1}(t)}{2}\sigma^{(1)}_{z}+\frac{\hbar \epsilon_{2} \omega_{2}(t)}{2}\sigma^{(2)}_{z}+\frac{\hbar \omega_{3}}{2}\sigma^{(1)}_{x}+\frac{\hbar \omega_{4}}{2}\sigma^{(2)}_{x}
+\frac{\hbar \epsilon_{3} \Omega_{c}(t)}{2}(\sigma^{(1)}_{x}\sigma^{(2)}_{x}+\frac{1}{30}\sigma^{(1)}_{z}\sigma^{(2)}_{z})
\end{equation}
with $\epsilon_{j}\in [0.8, 1.2]$ ($j=1,2,3$). Here, we assume that the frequencies $\omega_{1}(t), \omega_{2}(t) \in [-5, 5]\ \text{GHz}$ can be adjusted by changing the bias currents of the two phase qubits, and $\Omega_{c}(t) \in [-500, 500]\ \text{MHz}$ can be adjusted by changing the bias current in the coupler. Let $\omega_{3}=\omega_{4}=2\ \text{GHz}$, the operation time $T=20\ \text{ns}$ is divided into 40 smaller time intervals, and the step-size is 0.1. The initial control fields are $\omega_1(t)=\omega_2(t)=\sin t\ \text{GHz}$, $\Omega_c(t)=0.05\sin t \ \text{GHz}$.
Without fluctuations (i.e., $\epsilon_{j}\equiv 1$), the fidelity of the CNOT gate can reach $1-10^{-15}$ after 550 iterations, as shown in Fig. 7. When the uncertainty bounds are 0.2,
the results are shown in Fig. 8. In the training step, the precision of the CNOT gate can reach 0.9965. Then the average fidelity of 0.9961 can be reached for 2000 additional samples in the testing step.
\begin{figure}
\centering
\includegraphics[width=0.6\textwidth]{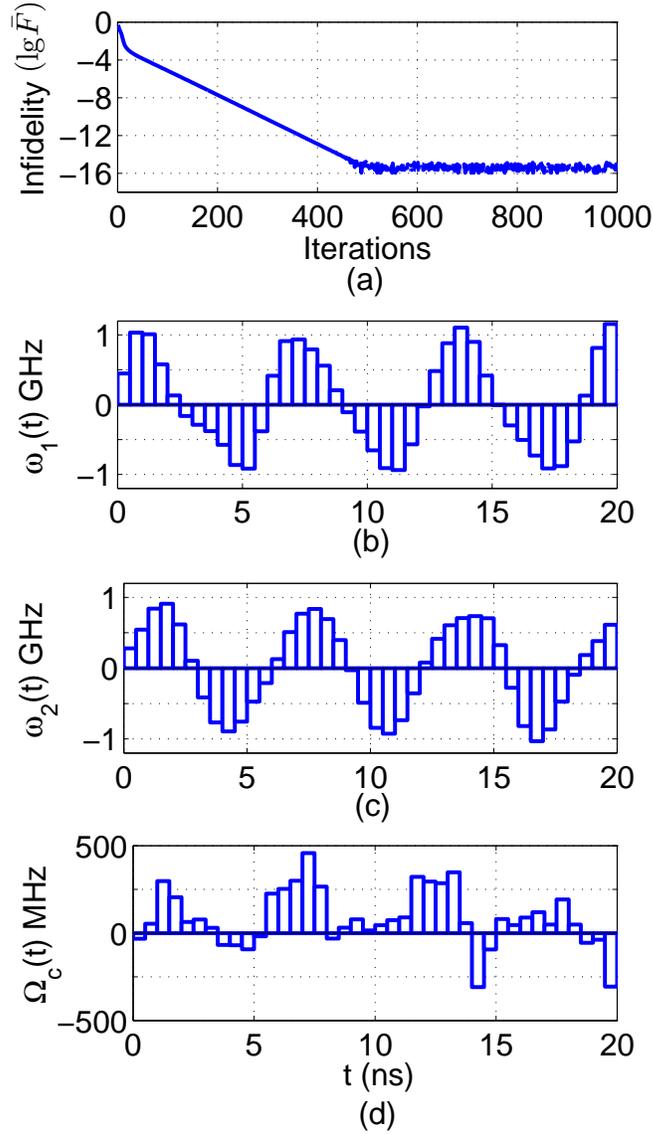}
\caption{The performance for constructing CNOT gates using optimal control fields. (a) Infidelity ($\lg\bar{F}$) versus iterations. (b), (c), (d) Learned optimal control fields.}\label{CNOT-optimal}
\end{figure}

\begin{figure}
\centering
\includegraphics[width=0.6\textwidth]{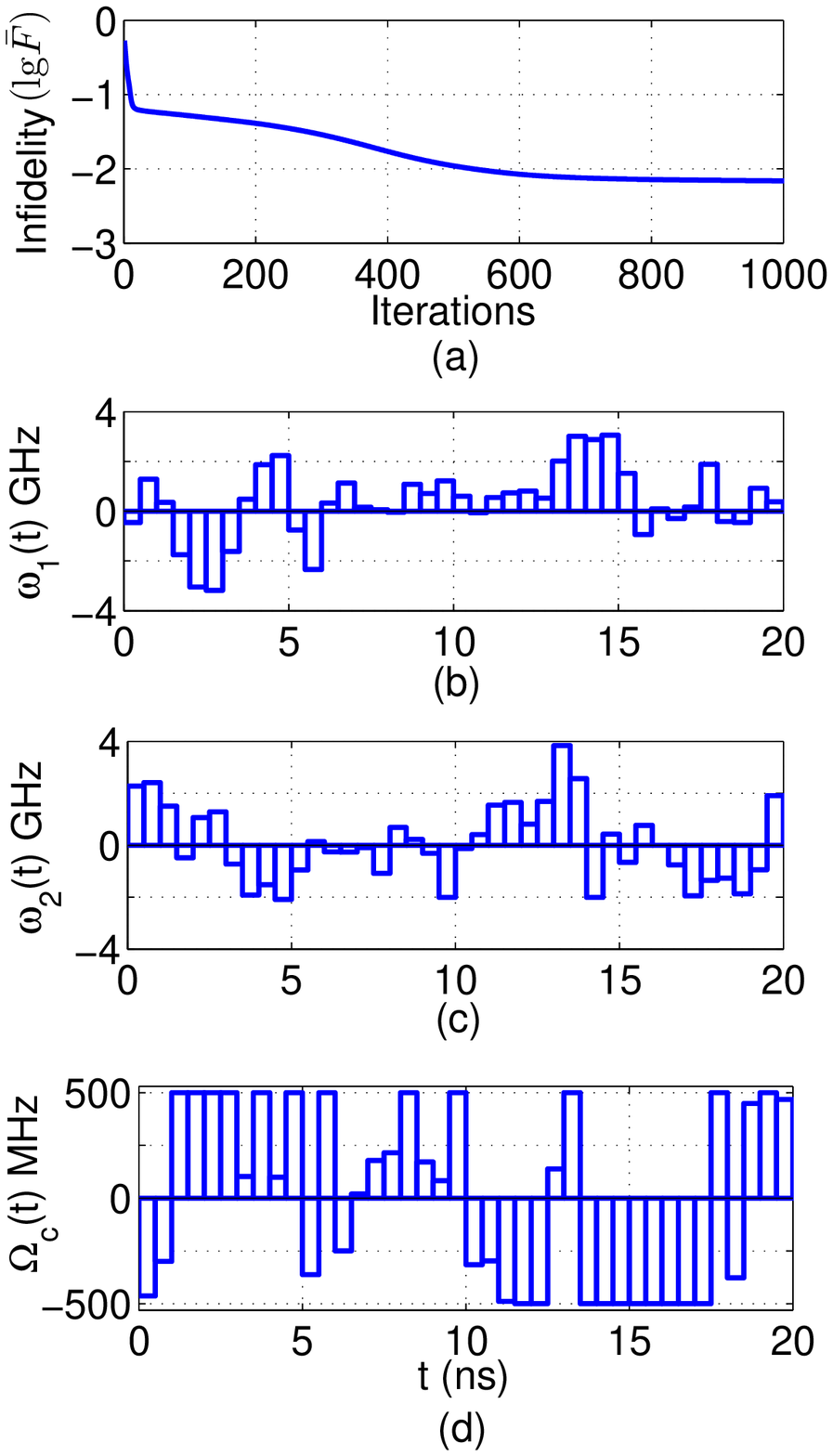}
\caption{Robust control performance for CNOT gates, with parameter fluctuations on $\omega_1$,  $\omega_2$ and $\Omega_c$, $E=0.2$. (a) Infidelity ($\lg\bar{F}$) versus iterations. (b), (c), (d) Learned robust control fields.}\label{CNOT-robust}
\end{figure}

\section*{Discussion}

In conclusion, we applied a learning-based open-loop method to find robust control fields for constructing a set of universal quantum gates. All of the quantum gates we achieved show good robustness and the method used is easy to implement.
Although a uniform distribution is used for the uncertainty parameters, the method is also applicable to other distributions (e.g., a Gaussian distribution). Several specific examples have been considered in this paper, while the method can be applied to other systems for constructing any quantum gates \cite{Ashhab-et-al-2012}. Also, it is straightforward to extend the method to achieving robust control tasks for general quantum operations.

\section*{Methods}

\subsection*{Gradient-based learning for quantum optimal control.} We first consider the unitary dynamics.
The evolution of a quantum gate $U(t)$ satisfies
\begin{equation}
  \dot{U}(t)=-iH(t)U(t), \ \ \ \ U(0)=I.\\
\end{equation}
Now the objective
is to design the Hamiltonian $H(t)$ to robustly steer the unitary $U(t)$ from $U(0)=I$ to the desired target
$U_{F}\in \{\mathrm{H}, \mathrm{S}, \mathrm{T_{\frac{\pi}{8}}}, \mathrm{CNOT}\}$, with high fidelity.
The fidelity is defined as:
\begin{equation}
F(U_F,U(T))=\frac{1}{2^{q}} |\langle U_F|U(T)\rangle|,                          
\end{equation}
where $q$ ($q=1$ or 2 in this paper) is the number of qubits involved in the quantum gate and $\langle U_F|U(T)\rangle\equiv\text{tr}(U_F^{\dag}U(T))$.

Gradient-based optimization algorithms have proven to be one of the most efficient methods
to solve optimization problems in quantum control. By applying a gradient-based optimization algorithm,
we now consider the problem of generating a high-fidelity quantum gate in a given time $T$.
Assume that the Hamiltonian has the following form
\begin{equation}
H(t)=H_0+\sum_{m=1}^{M}u_m(t)H_m,
\end{equation}
where $H_0$ is the free Hamiltonian, $H_m\ (m=1,2,\cdots M)$ are related to the control Hamiltonian with the corresponding control pulses $u_m(t)$.

The performance function of the transfer process can be defined as
\begin{equation}
\|U_F-U(T)\|^2=\|U_F\|^2-2\text{Re}\langle U_F|U(T)\rangle+\|U(T)\|^2.
\end{equation}
In practical applications, considering the possible existence of an arbitrary global phase factor $e^{i\varphi}$,
we minimize
\begin{equation}
\|U_F-e^{i\varphi}U(T)\|^2=\|U_F\|^2-2\text{Re}\langle U_F|e^{i\varphi}U(T)\rangle+\|e^{i\varphi}U(T)\|^2,
\end{equation}
which is equivalent to maximize $\text{Re}\langle U_F|e^{i\varphi}U(T)\rangle$.
In order to eliminate the influence of the global phase factor,
we maximize the performance function
\begin{equation}
\Phi=|{\langle U_F|e^{i\varphi}U(T)\rangle}|^2.
\end{equation}

Let $U_j$ denote the unitary transformation when the $j$th pulse $u^j$ is applied. We can decompose $U(T)$ as $U(T)=U_N\cdots U_1$.
With operators $A_j$ and $B_j$ being defined as $A_j=U_j\cdots U_1$ and
$B_j=U_{j+1}^\dagger \cdots U_N^\dagger U_F=A_jU(T)^{\dag} U_F$,
we have the following relationship
\begin{equation}
\Phi=\langle B_j|A_j\rangle \langle A_j|B_j\rangle.
\end{equation}
The gradient $\partial\Phi/\partial u_m(j)$ to the first order in $\Delta t$ is given by
\begin{eqnarray}
\frac{\partial\Phi}{\partial u_m(j)}&=&-\langle B_j|A_j\rangle \langle i\Delta t H_m A_j|B_j\rangle
-\langle B_j|i\Delta t H_m A_j\rangle \langle A_j|B_j\rangle
\nonumber\\
&=&-2\text{Re}\{\langle B_j|i\Delta t H_m A_j\rangle \langle A_j|B_j\rangle\}.
\end{eqnarray}
The optimal control field can be searched by following the gradient.

\subsection*{Open GRAPE.} For an open dissipative system, its dynamics can be described by a master equation. We will use an OPEN GRAPE algorithm to calculate the gradient (see \cite{OpenGrape4-Glaser-2011,OpenGrape2-Robentrost-2009,OpenGrape1-Goan-2012,OpenGrape3-Goan-2014}). We assume that the state of the system is described by a master equation
\begin{equation}
\dot{\rho}(t)=-(i\mathcal{H}(u(t))+\Lambda(u(t)))\rho(t)
\end{equation}
with the Hamiltonian superoperator $\mathcal{H}(u(t))(\cdot)=\frac{1}{\hbar}[H(u), \cdot]$, and the decoherence superoperator $\Lambda(u(t))$. The solution to the master equation is a linear map, according to $\rho(t)=\mathcal{G}(t)\rho(0)$. Hence, $\mathcal{G}(t)$ follows the operator equation
\begin{equation}
\dot{\mathcal{G}}(t)=-(i\mathcal{H}+\Lambda)\mathcal{G}(t)
\end{equation}
with $\mathcal{G}(0)=I$. The objective is to find a control field $u(t)$ to maximize the fidelity with a given final time $T$
\begin{equation}
F(U_{F}, \mathcal{G}(T))=\frac{1}{2^{q}}| \text{tr}\{U^{\dagger}_{F}\mathcal{G}(T)\}|
\end{equation}
The gradient of $F(U_{F}, \mathcal{G}(T))$ can be calculated using the method in \cite{OpenGrape2-Robentrost-2009} and the control field can be updated using the gradient.

\subsection*{Sampling-based learning control for robust design.}

The sampling-based learning control (SLC) method \cite{Chen-et-al-2013} involves two steps of training and testing. In the training step, we select $N$ samples to train the control fields. These samples are selected according to the distribution of uncertain parameters (e.g., uniform distribution). For example, when
 \begin{equation}
H(t)=\epsilon_{0}H_{0}+\sum_{m=1}^{M}\epsilon_{m}u_{m}(t)H_{m},
\end{equation}
an augmented system can be constructed as follows
\begin{equation}
\left(
\begin{array}{c}
  \dot{U}_{1}(t) \\
  \dot{U}_{2}(t) \\
  \vdots \\
  \dot{U}_{N}(t) \\
\end{array}
\right)= -i
\left(
\begin{array}{c}
  {H}^{s}_{1}(t){U}_{1}(t) \\
  {H}^{s}_{2}(t){U}_{2}(t) \\
  \vdots \\
  {H}^{s}_{N}(t){U}_{N}(t) \\
\end{array}
\right)\label{generalized-system}
\end{equation}
where the Hamiltonian of the $n$th sample $H^{s}_{n}(t)=\epsilon_{0n}H_{0}+\sum_{m=1}^{M}\epsilon_{mn}u_{m}(t)H_{m}$, with $n=1,2,\ldots,N$. The performance function of the augmented system is defined as
$F_N(u)$
\begin{equation}
F_N(u)=\frac{1}{N} \sum_{n=1}^{N} \frac{1}{2^{q}} |\langle U_F|U_{n}(T)\rangle|.
\end{equation}
The task of the training step is to find an optimal control $u^*$ which maximizes the performance function above.
The representative samples for these uncertain parameters can be selected according to the method in \cite{Dong-SR}.
In the testing step, we apply the control field $u^*$ we obtained in the training step to a large number of other additional samples, which are randomly selected according to the uncertainty parameters. If the average fidelity of all the tested samples is satisfactory, we accept the designed control, which means the quantum gate we constructed is robust.
In this paper, we use 2000 additional samples to test our designed control in this step. When the quantum system under consideration is an open system, uncertainties can exist in the decoherence parameters. For these uncertainty parameters, we can use a similar method for sampling these parameters to find robust control pulses.

\section*{Acknowledgment}
This work was supported by the Australian Research Council (DP130101658, FL110100020) and was
supported in part by the National Natural Science Foundation of
China (Nos. 61273327 and 61374092). F.N. was supported by a grant from the John Templeton Foundation, the RIKEN iTHES Project, MURI Center for Dynamic Magneto-Optics, and a Grant-in-Aid for Scientific Research (A).

\end{document}